\documentclass[pre,twocolumn,showpacs]{revtex4}

\usepackage{graphicx}
\usepackage{bm}

\newcommand{\avec}[1]{{\bm{#1}}}
\newcommand{\avecu}[1]{{\hat{\bm{#1}}}}

\begin{document}

\title{Simulational studies of axial granular segregation in a rotating cylinder}

\author{D. C. Rapaport}
\email{rapaport@mail.biu.ac.il}
\affiliation{Physics Department, Bar-Ilan University, Ramat-Gan 52900, Israel}

%\date{\today}
\date{February 25, 2002}

\begin{abstract}

Discrete particle simulation methods have been used to study axial segregation in a
horizontal rotating cylinder that is partially filled with a mixture of two
different kinds of granular particles. Under suitable conditions segregation was
found to occur, with the particles separating into a series of bands perpendicular
to the axis. In certain cases the band structure exhibited time-dependent behavior,
including band formation, merging and motion along the axis, all corresponding to
phenomena that arise experimentally. In order to examine how the many parameters
specifying the problem affect the segregation process, simulation runs were carried
out using a variety of parameter settings, including combinations of friction
coefficients not realizable experimentally. Both segregation and desegregation
(mixing) were investigated, and cylinders with both explicit end caps and periodic
ends were used to help isolate the causes of segregation.

\end{abstract}


\maketitle

\section{Introduction}

The properties of granular media are often very different from systems governed by
thermodynamics and statistical mechanics, and the intuition gained from these
theories proves to be of little help in trying to understand the mechanisms
responsible for granular behavior. The propensity of noncohesive granular mixtures
to segregate into individual species is one of the more conspicuous of these
properties, and the fact that segregation occurs even when then there is no
apparent energetic or entropic advantage is what makes such behavior so
fascinating. Since mixing and segregation are important processes, both in industry
and in nature, considerable effort has been invested in exploring the underlying
causes of these phenomena. In the absence of a general theory of granular matter,
much of the work in this field \cite{bar94,her95,jae96,kad99} has been, and
continues to be based on computer simulation.

Segregation of polydisperse granular mixtures occurs in a variety of situations,
including sheared flow \cite{wal83,hir97}, vibration \cite{ros87,gal96}, and
rotation \cite{can95,zik94}; in the most familiar form of rotation, involving a
horizontal cylinder rotating about its longitudinal axis, there are two modes of
segregation, one radial, the other axial. Although other forms of segregation have
been explored using discrete-particle simulations, the case of axial segregation in
a cylinder, with the exception of a very limited study \cite{sho98}, has not been a
subject of detailed investigation.

The typical experiment employs a transparent horizontal cylinder with a circular
cross section, partially filled with a mixture of two kinds of granular particles,
and rotating at a constant rate. While the most direct observations focus on the
composition of the visible surface layer \cite{zik94}, more elaborate studies
involve probing the interior using magnetic-resonance imaging (MRI) techniques
\cite{hil97s,hil97f}. Under suitable conditions the system is observed to segregate
into bands of alternating composition normal to the cylinder axis. Segregation, as
seen from the outside, may be so thorough that each band is composed exclusively of
a single particle species, with narrow transition regions separating the bands;
segregation may also be only partial, in the sense that the bands are characterized
by just a higher concentration of one species or the other. Beneath the surface,
segregation in the radial direction is also possible; in certain instances the
effects visible at the outer surface are in fact due to an axial core of one
species whose shape undulates along the axis, and the observed bands are a
consequence of this core intermittently extending to the outer surface
\cite{hil97f}. The band patterns can also exhibit time dependence; examples of such
behavior involve coarsening, in which narrower bands merge (sometimes very slowly)
to form broader patterns \cite{fre97}, and the appearance of traveling wave
patterns associated with band motion \cite{cho97,cho98}. The foregoing, partial
catalog of observations combines experimental results conducted under different
conditions with a variety of granular materials; there is currently no single set
of experiments providing systematic coverage of the dependence on the many
parameters contributing to the definition of the system.

The goal of the simulations reported in this paper is a systematic analysis of
certain aspects of the axial segregation problem, based on a molecular dynamics
approach involving discrete-particle models, making reference to experiment where
results are available. The model used here is of a kind routinely employed for
simulating granular flow phenomena. The degree to which such models prove capable
of reproducing the behavior of real granular media provides the ultimate validation
of this approach.

\section{Background}

Recent attempts at understanding axial segregation in a rotating horizontal
cylinder began with a series of experimental studies \cite{zik94} that considered
two types of mixtures, glass spheres and sand, and glass spheres of mixed sizes. In
experiments that started with a homogeneous binary mixture, bands were observed to
appear only for the glass-sand mixture, suggesting that different particle sizes
alone is not enough to produce segregation, but that different degrees of roughness
are required as well; it was also suggested that segregation required the rough
particles to be smaller than the smooth. Bands of various widths typically
developed after about 100 rotations; the narrowest bands subsequently disappeared,
leading eventually to a steady state characterized by bands with a narrow
distribution of widths. The importance of particle flow at the upper free surface
was established by noting that segregation ceased when this flow was obstructed.
Further conclusions were that segregation is not initiated at the cylinder ends but
is in fact driven by concentration fluctuations, and the importance of the fact
that the free surface has a characteristic curved shape rather than being flat.
Other experiments \cite{hil97f} also considered glass sphere mixtures of different
sizes and were able to achieve axial segregation; the discrepancy between these two
experiments, presumably a consequence of other differences in the experimental
conditions, has not been addressed. Further information concerning the early
experimental history of this phenomenon is to be found in Refs.~\cite{hil97f} and
\cite{sta98}.

The effects visible at the surface provide an incomplete description of the
behavior. Use of MRI \cite{hil97s,hil97f} facilitated examination of the interior,
in terms of both axial and radial cross sections, allowing the possible role of
subsurface dynamics to be examined. A mixture of plastic spheres and (smaller)
MRI-sensitive pharmaceutical pills was used in these experiments. Even when
segregation bands were observed at the surface, beneath the surface the small
particles were found to be present along the entire length of the cylinder, and the
appearance of bands of small particles at the surface merely corresponded to a
thickening of this interior region; once the presence of the undulating axial core
was discovered, band merging appears to be a less dramatic event than would
otherwise be the case. Another MRI study \cite{nak97}, in which both particle sizes
were visible, found the axial core of small particles to be a transient structure
that eventually disappeared, resulting in a state of complete axial segregation.
MRI has also been used in the study of the development of the interface between
initially segregated species \cite{rna99}.

Another aspect of axial segregation, namely the appearance of patterns that travel
along the cylinder in an axial direction, is described in Ref.~\cite{cho97,cho98},
using a mixture of sand and salt grains of different sizes. Only the behavior at
the outer surface (both the free surface and the exterior surface in contact with
the transparent cylinder wall) was examined, and a preferred wavelength was found
that was almost independent of rotation rate. In these experiments, the particle
sizes extended over broad, but nonoverlapping distributions, and the traveling
waves were found to occur only for certain volume fractions; furthermore, if the
salt grains were replaced by rounded sand particles of similar size no traveling
waves appeared. Details of various types of band motion and coarsening in a mixture
of sand and glass spheres are reported in Ref.~\cite{fre97}; the results suggest
that avalanches play an important role, and that the surface transport consists
mainly of the glass spheres moving across the sand, rather than the converse.
Extensions of the basic problem have also been considered, for example, cylinders
with axially modulated radius \cite{zik94}, and cylinders with noncircular
cross-section \cite{hil01}.

Given the complexity of the experimental situation it is hardly surprising that
theoretical progress has been limited. The experimental results in
Ref.~\cite{zik94} were accompanied by a simplified theory that dealt with the onset
of instability produced by local concentration fluctuations, and the response to
resulting changes in the angle of repose (as reflected, for example, in the
appearance of a bump where the granular surface meets the ascending wall of the
cylinder, and a depression at the opposite descending wall); homogeneity of the
cross-sectional composition within each band was assumed, something subsequent MRI
results failed to support. Continuum models have been developed based on
one-dimensional systems \cite{ara99a,ara99b} in which the dynamical variables are
the local concentration difference and the slope of the free surface; while such
models are able to describe the early phase of segregation with traveling bands and
subsequent band merging, important three-dimensional aspects of the problem are
absent and, contrary to experiment, the free surface profile is assumed linear. The
problem has also been studied using a simplified cellular automata model
\cite{yan99}, again assuming that the key to the behavior is what happens at the
free surface.

There is just one existing molecular dynamics (MD) study of axial segregation
\cite{sho98}. The system considered is small, only 1000 particles, and contained in
a relatively short cylinder, with behavior dominated by the end walls, and no
provision for rotation of the particles themselves. Partial segregation was
observed (the formation of two bands in which one of the species had a higher
concentration, separated by a single band dominated by the other), but only if the
bigger particles were the rougher ones, the opposite of the criterion in
Ref.~\cite{zik94}; different degrees of roughness but just a single particle size
failed to produce segregation. There is another, related MD study \cite{ris99} that
does not address segregation directly, but treats what amounts to the converse
problem, namely axial diffusion at the interface between initially segregated
species. Another kind of segregation can occur at comparatively low rotation rates,
this time radially, in which small particles tend to migrate inwards leaving a
higher concentration of big particles on the outside; this has been modeled in two
dimensions \cite{dur97}.

Despite the brevity of the above survey, it is abundantly clear that while
significant results have been obtained that help explain some aspects of axial
segregation, there is much that remains unresolved, both in regard to what actually
occurs in the system and the mechanisms underlying the phenomenon, a conclusion
also expressed in Ref.~\cite{sta98}. It has yet to be established whether there is
just a single kind of axial segregation phenomenon, or whether different kinds of
granular materials undergo segregation by invoking alternative mechanisms; it is
indeed possible that a number of processes operate concurrently, with the dominant
one being determined by the prevailing experimental conditions. It goes without
saying that the current state of experiment limits the ability to determine how
accurately simulation represents reality; further experimental evidence clearly
needs to be obtained.

\section{Granular model}

A variety of models have been employed in granular simulation, based on both soft
\cite{cun79,wal83,haf86} and hard \cite{cam90} particles; though referred to as
``soft'', this characterization applies more to the continuous nature of the
potential function used to oppose particle overlap during collision than to the
amount of overlap itself, which is minimal. Soft particles are used more widely
(including here) because of the simpler nature of the computations involved. In
addition to this excluded-volume repulsion force, the particles are also subject to
dissipative forces. The interactions fall into two classes, those between pairs of
particles, and those between the particles and the container walls, which, as will
be shown below, can be treated as special cases of particle-particle interactions.
The discussion of the model begins with the excluded-volume force.

Consider a pair of spherical granular particles $i$ and $j$ with diameters $d_i$
and $d_j$, respectively. The repulsive force that acts between particles closer
than a specified range \cite{hir01} is
\begin{equation}
\avec{f}_v = {48 \over r_{ij}} \left[ \left( {d_{ij} \over r_{ij}} \right)^{12} -
{1 \over 2} \left( {d_{ij} \over r_{ij}} \right)^6 \right] \avecu{r}_{ij} ,
\label{eq:fv}
\end{equation}
where $\avec{r}_{ij} = \avec{r}_i - \avec{r}_j$ is the particle separation, $r_{ij}
= |\avec{r}_{ij}|$, and the effective diameter entering the interaction
computations is $d_{ij} = (d_i + d_j) / 2$. This force is derived from the
functional form of the Lennard-Jones potential; it acts whenever $r_{ij} < 2^{1/6}
d_{ij}$ and is continuous at the cutoff point (although its derivative is not).
Alternatives to this form of overlap interaction that are also in routine use
include functions that depend on the overlap either linearly or to the 3/2-power
\cite{sch96}, but the choice of function is of little consequence as far as the
bulk behavior is concerned \cite{rap01}. Note that because of the slight degree of
softness, particle diameter is not precisely defined.

What distinguishes the interactions used for granular media from those in MD
studies of molecular systems is the presence of dissipative forces that act over
the duration of each collision. The first of these is a normal (viscous) damping
force
\begin{equation}
\avec{f}_d = - \gamma_n (\avecu{r}_{ij} \cdot \avec{v}_{ij}) \avecu{r}_{ij},
\label{eq:fd}
\end{equation}
that depends on the component of the relative velocity of the particles
$\avec{v}_{ij} = \avec{v}_i - \avec{v}_j$ in the direction between the particle
centers. The factor $\gamma_n$ is the normal damping coefficient, assumed to be the
same for all particles. The total force parallel to $\avec{r}_{ij}$ is $\avec{f}_n
= \avec{f}_v + \avec{f}_d$.

Frictional damping also acts in the transverse direction at the point of contact of
the particles. The relative transverse velocity of the particle surfaces at this
point, allowing for particle rotation, is
\begin{equation}
\avec{v}_{ij}^s = \avec{v}_{ij} - (\avecu{r}_{ij} \cdot \avec{v}_{ij})
\avecu{r}_{ij} - \left( {d_i \avec{\omega}_i + d_j \avec{\omega}_j \over
d_i + d_j} \right) \times \avec{r}_{ij} ,
\end{equation}
where $\avec{\omega}_i$ is the angular velocity of particle $i$. The sliding
friction is then
\begin{equation}
\avec{f}_s = - \min \left( \gamma_s^{c_i c_j} |\avec{v}_{ij}^s|, \,
\mu^{c_i c_j} |\avec{f}_n| \right) \avecu{v}_{ij}^s .
\label{eq:ft}
\end{equation}
Here, $\gamma_s^{c_i c_j}$ is the sliding friction coefficient, whose value depends
on the species $c_i$ and $c_j$ of the particles involved. The static friction
coefficient $\mu^{c_i c_j}$ sets an upper bound to the sliding friction
proportional to $|\avec{f}_n|$; in models of this kind there is no true static
friction (at best, e.g., Ref.~\cite{dur97}, it can be represented by a tangential
restoring force that depends on the relative displacement occurring during contact,
although it should be emphasized that this is not a strictly correct means for
incorporating the effects of static friction).

The translational and rotational accelerations of the particles, $\avec{a}_i$ and
$\avec{\alpha}_i$ depend on sums of the above terms for all interacting pairs. Each
such contribution is included by adding $\avec{f}_n + \avec{f}_s$ to the total $m_i
\avec{a}_i$, subtracting the same quantity from $m_j \avec{a}_j$, and also
subtracting  $\avecu{r}_{ij} \times \avec{f}_s / \kappa$ from both $m_i d_i
\avec{\alpha}_i$ and $m_j d_j \avec{\alpha}_j$, where $m_i = d_i^3$ is the particle
mass and $\kappa$ the numerical factor in the moment of inertia (for a solid sphere
$\kappa = 0.2$).

Similar considerations apply to the interactions between particles and container
walls. While there are alternative methods for representing boundaries, for
example, by constructing explicit rough boundaries out of (a large number of)
constrained particles with properties similar to the mobile particles, or by
treating them separately from the interactions by simply moving particles back
inside if they are found to have crossed a boundary, the approach used here is
based on walls represented by geometrically smooth surfaces with similar frictional
properties to the particles themselves.

The curved cylinder wall is treated as follows, assuming the axis of the cylinder
to be in the $y$-direction. The value of $d$ associated with the wall is unity, so
that $d_{iw} = (d_i + 1) / 2$ replaces $d_{ij}$ in the force calculations. To
determine whether a particle lies within interaction range of the curved boundary
evaluate $\avec{r}_{iw} = \avec{r}_i - (\avec{r}_i \cdot \avecu{y}) \avecu{y}$; the
particle is in range if $r_{iw} > D / 2 - 2^{1/6} d_{iw}$, where $D$ is the
cylinder diameter. To evaluate the forces replace $\avec{r}_{iw}$ by
$\avec{r}'_{iw} = (1 - D / 2 r_{iw}) \avec{r}_{iw}$ and evaluate $\avec{f}_v$ using
Eq.~(\ref{eq:fv}). The velocity of the cylinder wall at the effective point of
contact is $\Omega \avec{s}$, where $\Omega$ is the cylinder angular velocity and
$\avec{s} = (D / 2) \avecu{r}_{iw} \times \avecu{y}$ is parallel to the direction
of motion of the cylinder wall at the point of contact; thus the relative velocity
corresponding to $\avec{v}_{ij}$ is $\avec{v}_{iw} = \avec{v}_i - \Omega \avec{s}$.
Evaluate $\avec{f}_d$ using Eq.~(\ref{eq:fd}); the resulting $\avec{f}_n$
contributes to $\avec{a}_i$. Evaluate the sliding velocity at the contact point,
$\avec{v}_{iw}^s = \avec{v}_{iw} - (\avecu{r}'_{iw} \cdot \avec{v}_{iw})
\avecu{r}'_{iw} - \avec{\omega_i} \times \avec{r}'_{iw}$, and use this in an
expression similar to Eq.~(\ref{eq:ft}) to compute the sliding friction,
\begin{equation}
\avec{f}_s = - \min \left( \gamma_s^{c_i w} |\avec{v}_{iw}^s|, \,
\mu^{c_i w} |\avec{f}_n| \right) \avecu{v}_{iw}^s ,
\end{equation}
which contributes to both $\avec{a}_i$ and $\avec{\alpha}_i$.

The cylinder ends can be periodic, if the goal is to avoid any spurious wall
effects, or capped, using walls with the same frictional properties as the curved
cylinder wall. In the latter case the wall contributions are computed as follows.
First check whether the particle is in range of either of the end walls by
evaluating $\avec{r}_{iw} = (|\avec{r}_i \cdot \avecu{y}| - L / 2) \avecu{y}$,
where $L$ is the cylinder length; if $\avec{r}_{iw} \cdot \avecu{y} > - 2^{1/6}
d_{iw}$, where $d_{iw}$ was defined above, then the particle is within range. In
that case the calculation is as before, except that $\avec{r}'_{iw} = \pm
\avec{r}_{iw}$, whichever points towards the origin, and $\avec{s} = \avec{r}_i
\times \avecu{y}$. Finally, the effect of gravity is included by subtracting $g
\avecu{z}$ from each $\avec{a}_i$, where $g$ is the gravitational acceleration.

Several friction coefficients appear in the model; some are fixed at a single
value, while others are assigned a choice of values to examine their effect on the
behavior. The normal damping coefficient $\gamma_n$ is an example of the former,
whereas the particle-particle (P-P) sliding friction coefficients for identical
particles $\gamma_s^{bb}$ and $\gamma_s^{ss}$, where $b$ and $s$ denote big and
small particles, vary between runs, as do the particle-wall (P-W) coefficients
$\gamma_s^{bw}$ and $\gamma_s^{sw}$. For collisions involving mixed particle types,
$\gamma_s^{bs} = \min (\gamma_s^{bb}, \gamma_s^{ss})$ (an arithmetic or geometric
average could also have been used). The relative values of the static friction
coefficients $\mu^{bb} / \mu^{ss}$ and $\mu^{bw} / \mu^{sw}$ are set equal to the
ratio of the corresponding $\gamma_s$ values, with the larger of each pair having
the fixed value 0.5. While the fact that the P-W friction coefficients are
specified independently of the P-P coefficients allows situations that might be
experimentally unrealizable, such flexibility might nevertheless contribute to
learning more about the segregation mechanisms.

Other details of the simulation follow standard molecular dynamics procedure
\cite{rap95}. Neighbor lists are used to organize the force computations
efficiently. The translational and rotational equations of motion are integrated
using the leapfrog method (there is no need to evaluate the rotational coordinates
of the particles). Because of the heavy computations involved in some of the runs,
parallel computing methods based on a spatial decomposition of the system were used
to spread the workload across several coupled processors.

\section{Simulation parameters}

A substantial number of parameters are involved in specifying the system, and even
after fixing some of them quite a few remain; in order to reduce the number of runs
for different parameter combinations, typically just two values are considered
for each parameter. What occurs for other parameter combinations will not be
examined, but even the present selection offers a rich variety of phenomena; a more
complete analysis of the multidimensional phase diagram will require substantial
additional effort. Overall, there is a certain arbitrariness in many of the
parameter settings, although well-separated pairs of values have been used; some
parameters are of course constrained by the nature of the model and the demands of
computational stability. Parameter combinations that hinted at more interesting
behavior were studied in greater detail using larger systems and longer runs; with
so many choices available, this is a reasonable criterion for an initial
exploratory study.

The reduced units employed here are readily related to the corresponding physical
units. If $L_{MD}$ is the length unit (in m), then the corresponding time unit is
$T_{MD} \approx \sqrt{g L_{MD} / 9.8}$ s. For the value $g = 5$ used here, the time
unit becomes $T_{MD} \approx 10^{-2} \sqrt{5 L_{MD}}$, where, for convenience,
$L_{MD}$ is now measured in mm. The real rotation frequency is $\Omega / (2 \pi
T_{MD}) \approx 7.1 \Omega / \sqrt{L_{MD}}$ Hz, so for 3 mm particles, the
frequency in run \#A would be 2 Hz, a reasonable experimental value. In addition,
reduced units are implicit in Eq.~(\ref{eq:fv}) and in the definitions of the
friction coefficients.

Most of the runs involve mixtures of particles with two diameters, nominally 1 and
1.3 (in reduced units), but a few treat particles of the same size. A size ratio of
1.3 is smaller than the ratios typically employed in experiment, usually in the
range 2--4; the reason for this choice is computational convenience, since for a
larger size ratio the number of particles in the simulation would have to be
increased accordingly. In most cases the cylinder angular velocity $\Omega$ is 0.5
(radian/unit-time), but a lower value of 0.1 is also used.  Several values of
cylinder length $L$ and diameter $D$ are included, with an aspect ratio (a number
sometimes quoted in the experiments) ranging between 8:1 and 20:1. The initial
state of each run consists of a cylinder uniformly filled with particles arranged
as an FCC lattice with a specified number density $\rho$; varying $\rho$ allows the
fill level of the cylinder to be adjusted. The initial particle velocities are
random.

Several parameters are assigned a single value: the normal damping coefficient
$\gamma_n = 5$, gravitational acceleration $g = 5$, equal numbers of particles of
either species, and a uniform random distribution of particle diameters in the
narrow range $[d - 0.2, d]$, where $d$ is the nominal diameter of the species. The
integration timestep is $\delta t = 5 \times 10^{-3}$ (in reduced units).

The initial state is either mixed or segregated, and the cylinder ends periodic or
capped. The precise number of particles in the system (which determines the fill
level for a given cylinder size) is determined by the packing density of the
initial state $\rho$ (whose value in most cases is 0.3), and to a lesser degree by
the boundary conditions and the nature of the initial state. The total number of
particles $N$ in the runs reported here ranges from approximately 3,300 to 62,200,
while the number of cylinder rotations covered by the simulation ranges from 500,
adequate in some cases to establish that segregation (or mixing) occurs, to as many
as 38,000 required for examining slow pattern evolution (since the number of
integration timesteps per rotation is $2 \pi / \Omega \delta t$ certain runs are
quite long).

The selection of simulation runs described in the following section are listed in
Table~\ref{tab:runs}; the letter codes allow convenient referencing of the runs
without the need to repeat the actual parameter settings. A variety of sliding
friction coefficients are considered; the four separate $\gamma_s$ coefficients are
assigned values 2 or 10, but in different combinations, with no connection between
the P-P and P-W values; the actual value combinations used appear in the table.

\begingroup
\squeezetable
\begin{table}
\caption{\label{tab:runs}Summary or run parameters and other details; only some of
the runs appear in the figures.}
\begin{ruledtabular}
\begin{tabular}{lrrrrrrccrrrrrl}
Id\footnote{Runs are denoted in the text as, e.g., \#A; runs shown in figures are
starred.} & \multicolumn{4}{c}{Size\footnote{Cylinder length $L$ and diameter $D$,
initial filling density $\rho$, total number of particles $N$, big particle size
$b$, angular velocity $\Omega$.}} &&& Bdy\footnote{Cylinder end boundaries:
periodic or capped.} & Init\footnote{Initial state: mixed or segregated.} &
\multicolumn{4}{c}{Friction\footnote{Values of $\gamma_s$ for P-P and P-W
interactions, for big and small particles.}} & Turns & Bands\footnote{Number of
bands formed, if any (the space-time plots may lack the resolution to show rapidly
changing transients, and not all transient features are regarded as bands).} \\
&\hfil $L$ \hfil& $D$ & $\rho$ &\hfil $N$ \hfil&\hfil $b$ \hfil& $\Omega$ &
P,C & M,S & bb & ss & bw & sw & $\times 10^3$ & \\
\hline
A$^\ast$ &128&16& .3 & 4664  &1.3& .5 &P&M& 10&2 &10&2 &  4.3 & 8$\to$6   \\
B$^\ast$ &128&16& .3 & 4664  &1.3& .5 &P&M& 2 &10&10&2 &  5.8 & 10$\to$6  \\
C        &128&16& .3 & 4664  &1.3& .5 &P&M& 2 &2 &10&2 &  0.5 & 8         \\
D        &128&16& .3 & 4664  &1.3& .5 &P&M& 10&10&10&2 &  1.5 & 6         \\
E        &128&16& .3 & 4664  &1.3& .5 &P&M& 10&2 &10&10&  1.1 & 8         \\
F        &128&16& .3 & 4664  &1.3& .5 &P&M& 2 &10&2 &10&  0.5 & -         \\
G        &128&16& .3 & 4664  &1.3& .5 &P&M& 2 &2 &2 &10&  0.6 & -         \\
H        &128&16& .3 & 4664  &1.3& .5 &P&M& 10&2 &2 &10&  1.0 & -         \\
I$^\ast$ &256&16& .3 & 9416  &1.3& .5 &P&M& 10&2 &10&2 & 11.0 & 16$\to$8  \\
J$^\ast$ &256&24& .3 & 27392 &1.3& .5 &P&M& 10&2 &10&2 & 12.7 & 14$\to$6  \\
K$^\ast$ &256&24& .3 & 27392 &1.3& .1 &P&M& 10&2 &10&2 &  4.0 & 12$\to$6  \\
L$^\ast$ &128&16& .3 & 4664  &1.0& .5 &P&M& 10&2 &10&2 &  1.6 & 8$\to$6   \\
M        &128&16& .3 & 4664  &1.0& .5 &P&M& 2 &2 &10&2 &  3.5 & 2         \\
N        &128&16& .3 & 4664  &1.0& .5 &P&M& 10&2 &10&10&  3.8 & 2         \\
P$^\ast$ &128&16& .4 & 7080  &1.3& .5 &P&M& 10&2 &10&2 &  8.1 & 8$\to$6   \\
Q        &128&16& .5 & 9828  &1.3& .5 &P&M& 10&2 &10&2 &  5.9 & 6         \\
R$^\ast$ &128&16& .3 & 4576  &1.3& .5 &C&M& 10&2 &10&2 &  1.7 & 11$\to$9  \\
S$^\ast$ &320&16& .3 & 11704 &1.3& .5 &C&M& 10&2 &10&2 & 15.8 & 19$\to$11 \\
T        &320&32& .3 & 62244 &1.3& .5 &C&M& 10&2 &10&2 &  2.5 & 13$\to$9  \\
U$^\ast$ &320&16& .3 & 11704 &1.3& .1 &C&M& 10&2 &10&2 &  3.6 & 19        \\
V$^\ast$ &128&16& .3 & 3390  &1.3& .5 &P&S& 10&2 &10&2 & 11.1 & 2         \\
W        &128&16& .3 & 3390  &1.3& .5 &P&S& 10&10&10&10&  3.9 & -         \\
X        &128&16& .3 & 3346  &1.3& .5 &C&S& 10&2 &10&2 &  5.4 & 2         \\
Y$^\ast$ &128&16& .3 & 3346  &1.3& .5 &C&S& 10&2 &10&10& 38.0 & 2--6      \\
\end{tabular}
\end{ruledtabular}
\end{table}
\endgroup

\section{Analysis of results}

\subsection{Segregation in periodic containers}

One of the advantages of the simulational approach is that it permits the
realization of systems that are not readily constructed in the laboratory. In the
present case, the use of periodic boundaries rather than hard caps at the cylinder
ends can eliminate the possibility that segregation is due to end effects; if
segregation is indeed a consequence of fluctuations that can appear anywhere, then
axial periodicity provides the means of establishing this fact. Similarly, the
available choices of friction coefficients provide flexibility unavailable
experimentally.

\begin{figure}
\includegraphics[scale=0.65]{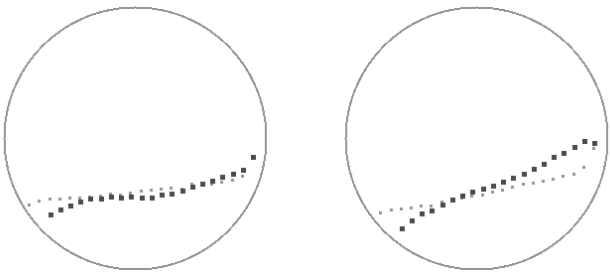}
\caption{\label{fig:pltjk}Axially averaged surface heights for runs \#K and \#J;
symbol size indicates particle size.}
\end{figure}

The first system considered here, \#A in Table~\ref{tab:runs}, has $L = 128$, $D =
16$, and a total of $N = 4664$ particles; the aspect ratio $L / D = 8$. The sloping
free surface that rapidly forms crosses the centerline of the tube approximately
midway between the axis and the boundary (surface profile plots are shown later, in
Fig.~\ref{fig:pltjk}, for similar but larger systems). The ratio of big and small
particle diameters is $b = 1.3$, the angular velocity is $\Omega = 0.5$, the
cylinder ends are periodic, and initially the two species are randomly mixed. The
sliding friction coefficients are $\gamma_s^{bb} = \gamma_s^{bw} = 10$,
$\gamma_s^{ss} = \gamma_s^{sw} = 2$, which means that the big particles are rougher
than the small, both in their interactions with each other and with the curved
cylinder wall. All this information is contained in Table~\ref{tab:runs} and will
not be detailed in subsequent cases.

A condensed summary of the observed behavior over the entire run is most
conveniently presented using a ``space-time'' plot showing local relative
concentrations of the two species, in which the horizontal scale measures elapsed
time and the vertical scale shows the position along the cylinder axis. The plot is
an accumulation of snapshots taken at regular intervals (roughly every 80th
rotation for this run); each such snapshot is represented by a narrow vertical
strip in the figure and shows the relative concentration (weighted by particle
volume) in a series of thin slices perpendicular to the rotation axis, with the
shading indicating the actual value, ranging from dark for a pure band of small
particles, to light for a band of big particles; gray values (not many appear here
because the results focus on cases with successful segregation) indicate mixtures
(without revealing any information about what occurs in the radial direction). All
plots are of the same size, regardless of run duration and cylinder length; the
size of a single ``pixel'' in each plot is inversely proportional to these two
quantities. The space-time plots used here, though visually similar to those in
experimental studies \cite{hil97f,fre97,cho98}, do not provide exactly the same
information as the latter, which are obtained from video signals that record
concentrations in the upper free surface of the material. In the case of
near-complete axial segregation, both convey the same information, but if the
particle distribution in the interior differs from the surface the images will have
very different meanings.

\begin{figure}
\includegraphics[scale=1.1]{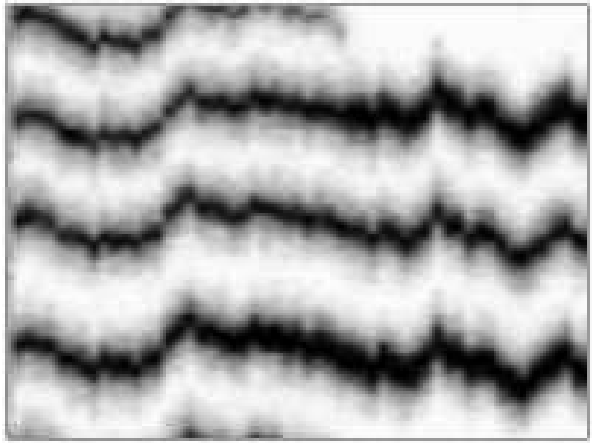}
\caption{\label{fig:sta}Space-time plot for run \#A; time advances horizontally
from left to right (the range is the entire duration of the run), and the vertical
scale corresponds to the axial position in the cylinder; dark shading is used for
bands of small particles (in the runs that involve mixed sizes).}
\end{figure}

The space-time plot for run \#A, which extends over 4300 cylinder rotations, is
shown in Fig.~\ref{fig:sta}. Eight bands, four each of big and small particles,
appear very shortly after the start of the run, just as in experiment, within the
first 50 rotations. Note that due to periodic wraparound, bands extend across the
top and bottom ends of the plot. Midway through the run one of the dark
(small-particle) bands disappears and the adjacent light bands merge. The resulting
pattern persists for the remainder of the run. While the spacing of the bands tends
to be fairly constant (aside from the effect of merging), the band pattern as a
whole exhibits apparently random oscillatory motion in the axial direction,
although the fluctuations tend to be less than the spacing between alternate bands.
The immediate conclusion from Fig.~\ref{fig:sta} is that the simulations are indeed
capable of producing axial segregation, and since the boundaries are periodic the
effect is a result of spontaneous symmetry breaking in the initially homogeneous,
axial direction (rather than being initiated by processes at the cylinder ends).

\begin{figure}
\includegraphics[scale=0.58]{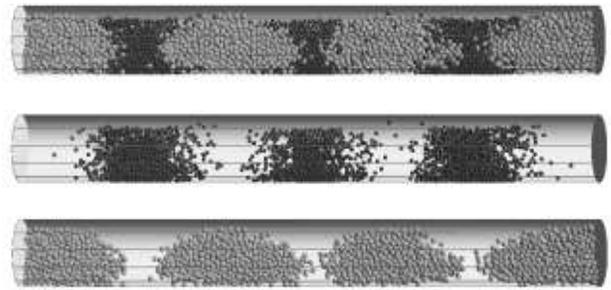}
\caption{\label{fig:vua}Images from run \#A: a view of the entire system seen from
above, followed by views of the small (darkly shaded) and big particles separately
(the narrow stripes seen here and in similar pictures are part of the container,
whose front surface is drawn as an open cage).}
\end{figure}

Since space-time plots provide only partial characterization of the behavior, more
detailed information is extracted by examination of actual configurations occurring
during the run. Figure~\ref{fig:vua} shows three simultaneous views of \#A, as seen
from above, at an angle approximately normal to the sloping upper surface. The
first image corresponds to what would be seen experimentally; it is followed by two
images showing the small and big particles separately (analogous to MRI imaging).
These images are taken from an animated sequence that allows detailed examination
of the particle motion. Segregation is now seen to be essentially complete, even at
the band centers, a result reminiscent of experiment \cite{nak97}.

\begin{figure}
\includegraphics[scale=1.1]{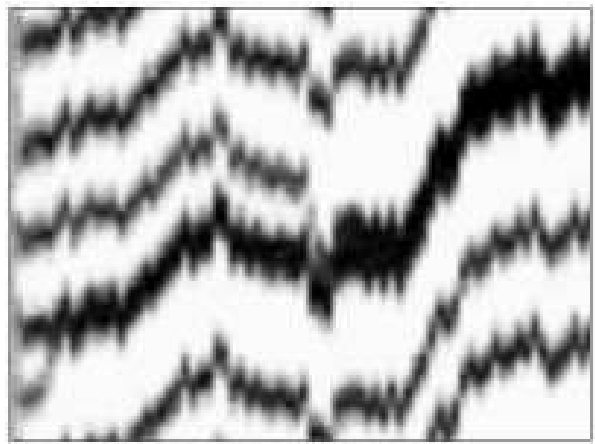}
\caption{\label{fig:stb}Space-time plot (\#B).}
\end{figure}

Run \#B, shown in Fig.~\ref{fig:stb}, treats a system similar to \#A but with the
wall friction coefficients reversed, so that the rougher, big particles experience
a smoother wall and vice versa. This run begins by forming ten bands, which
quickly merge to eight, and midway through the run an additional merger leaves
just six bands. The dark bands appear somewhat sharper here; this reflects the
fact that the boundaries between species are better defined than before, with even
fewer small particles to be found in among the big particles.

A series of additional runs \#C--\#H, for systems similar to \#A but with different
choices of sliding friction coefficients (in which there is no relation between the
P-P and P-W coefficients) leads to the following observations: if $\gamma_s^{bw} >
\gamma_s^{sw}$, segregation occurs irrespective of the relation between
$\gamma_s^{bb}$ and $\gamma_s^{ss}$; if $\gamma_s^{bw} = \gamma_s^{sw}$,
segregation requires $\gamma_s^{bb} > \gamma_s^{ss}$; if $\gamma_s^{bw} <
\gamma_s^{sw}$, there is no segregation. While these observations apply only to the
parameter sets examined, they do suggest that if the cylinder wall exerts a
stronger drag on the big particles, thereby raising their preferred surface slope
relative to the small, then segregation occurs, but not in the converse case. The
result from the earlier MD study \cite{sho98} is consistent with this statement. If
the wall affects both sizes of particle equally, so that the difference in the
surface profiles of the species is limited to a slightly altered shape rather than
any marked change in the overall slope, then segregation is only seen if the big
particles are rougher.

\begin{figure}
\includegraphics[scale=1.1]{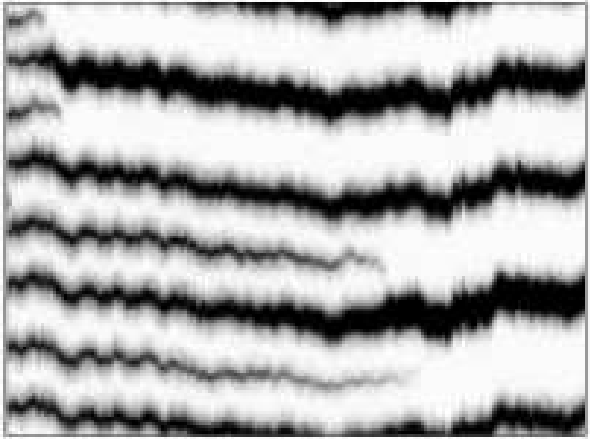}
\caption{\label{fig:sti}Space-time plot (\#I).}
\end{figure}

\begin{figure}
\includegraphics[scale=1.1]{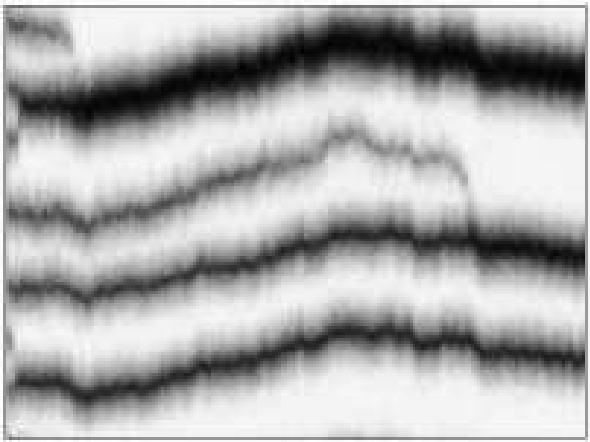}
\caption{\label{fig:stj}Space-time plot (\#J).}
\end{figure}

\begin{figure}
\includegraphics[scale=1.2]{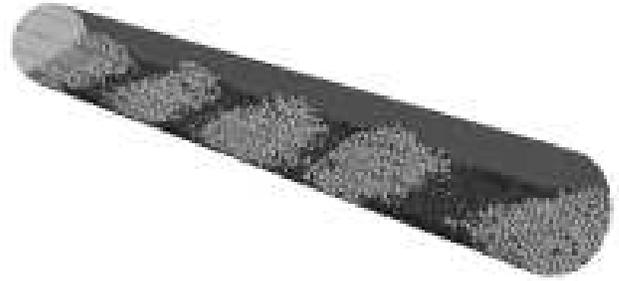}
\caption{\label{fig:vuj}Perspective view of run \#J showing surface heaping
(the cylinder ends are periodic and rotation is in a counterclockwise direction).}
\end{figure}

The systems studied so far have been fairly small; by considering larger systems it
is possible to encounter richer, more interesting behavior, although without any
change in the conclusions as to the dependence on the relative friction values. The
first case \#I, shown in Fig.~\ref{fig:sti}, is similar to \#A but with a longer
cylinder; here the initial 16 bands gradually contract to eight, with the changes
occurring at various times. The second example \#J in Fig.~\ref{fig:stj} involves a
wider cylinder, where, after merging, only six bands remain. Here the bands are
less sharply defined, hinting at the possibility of interesting behavior within;
when viewed from outside (both the upper surface and the material in contact with
the transparent cylinder wall) the bands suggest complete segregation, with just a
few particles in the wrong regions, but examination of the interior reveals that
while small particles are mostly confined to their bands, big particles are
separated by at best narrow gaps (see also run \#K below). The tendency of the big
particle region not to fragment is confined to an axial core that is invisible from
outside; this is reminiscent of behavior noted in studies using MRI (although
experimentally \cite{hil97s,hil97f} it is the MRI-visible small particles that
appear to concentrate near the axis). Figure~\ref{fig:vuj} is an oblique view of
the full system, with a container whose forward faces have been removed; heaping of
the rougher big particles at the upward moving rear boundary is clearly visible.

\begin{figure}
\includegraphics[scale=1.1]{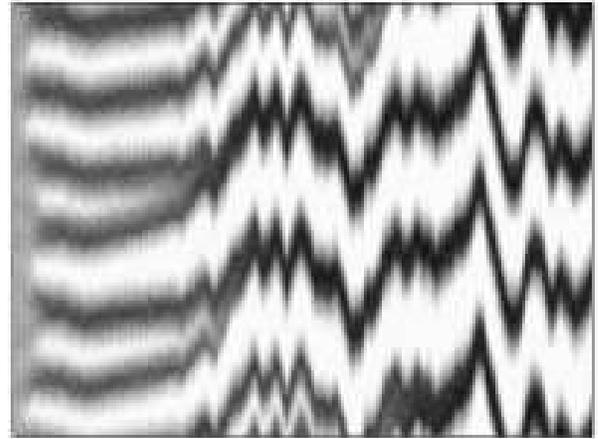}
\caption{\label{fig:stk}Space-time plot (\#K).}
\end{figure}

\begin{figure}
\includegraphics[scale=0.58]{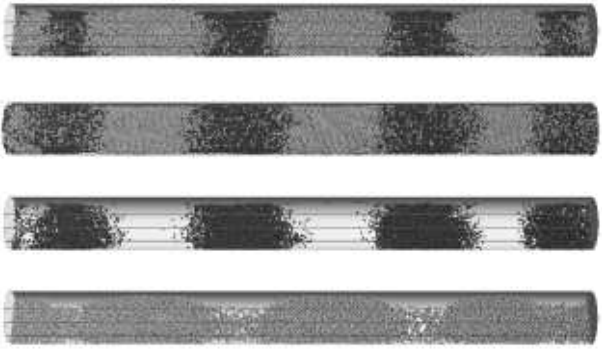}
\caption{\label{fig:vuk}Views of run \#K from above and below showing entire
system, and from above showing small and big particles separately.}
\end{figure}

The rotation rate is another of the parameters that can affect behavior; surface
slopes will be changed, and that in itself is enough to alter the particle flow,
but in addition the motion in the interior could be different. The system in
Fig.~\ref{fig:stk}, run \#K, corresponds to a reduced $\Omega$, but is otherwise
identical to \#J (but with fewer rotations). The initial uniformly gray region
represents the mixed state, out of which the bands emerge at similar, although not
identical times. Figure~\ref{fig:vuk} shows four simultaneous views of the system,
first as seen from above and below, corresponding to what would be visible in a
typical experiment, and then images of just the small and big particles that
provide some indication of the interior organization; the ability to look inside
reveals that there is a clear asymmetry between big and small particles (more so
than in run \#A), with the big particles appearing throughout the core region,
although on the outside there is little hint of this behavior. 

The cylinder diameter in runs \#J and \#K is sufficiently large for much of the
free surface to be reasonably far away from the curved wall, allowing meaningful
examination of the surface profiles (profiles for smaller diameters are more
strongly affected by wall proximity). Figure~\ref{fig:pltjk} shows plots of the
axially averaged surface heights across the system (each is averaged over 10
configurations separated by several turns) towards the ends of the runs. The
profiles in the two cases are very different; for lower $\Omega$ (\#K) the
principal difference between the big and small particles (the former are rougher)
appears near the boundaries, whereas for higher $\Omega$ (\#J) there is a
significant difference across the entire profile. In each case the detailed surface
shape is too complex (they are neither planar nor even ``S''-shaped) to be
characterized just by a dynamic angle of repose; furthermore, since the results
represent axial averages they do not reveal surface variation along this direction,
as can be seen in Fig.~\ref{fig:vuj}.

Rough estimates of the angles associated with these slopes (based on a visual
straight-line fit) lie in the approximate range $10^\circ$--$20^\circ$, less than
typical granular media where, for example, values in the range
$30^\circ$--$36^\circ$ are found \cite{zik94}. This is just one consequence of the
absence of static friction, the other is that the static angle of repose is
essentially zero, so that if rotation stops the surface returns to the horizontal,
a situation very different from experiment; on the other hand, the frictional
forces that are present in the model do produce a slope adequate for achieving
segregation, so that static friction may not play an essential role here.

\begin{figure}
\includegraphics[scale=1.1]{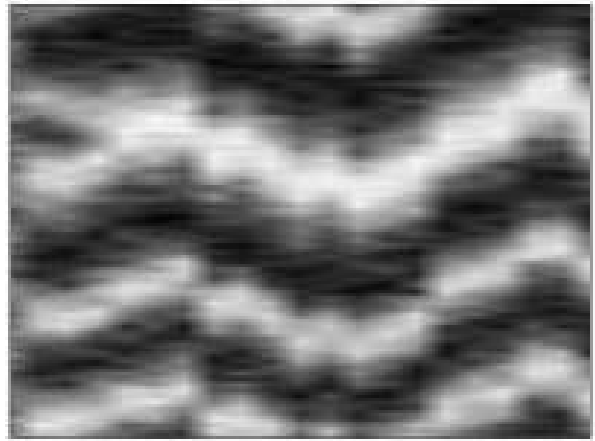}
\caption{\label{fig:stl}Space-time plot (\#L).}
\end{figure}

In the model under study, segregation is also found to occur for particles of
identical size, provided the frictional properties differ in a particular way. Run
\#L, in Fig.~\ref{fig:stl}, shows an example of this behavior; here the particles
labeled big and small have the same nominal diameter, but the other details are
identical to \#A (the shading just distinguishes the species). The bands are not as
sharp as before, but segregation still occurs, even if it no longer results in
total exclusion of each species from the bands of the other. This observation, with
further results from runs \#M and \#N as to which sets of friction coefficients
lead to segregation, is consistent with the earlier observations based on particles
of different sizes; if one particle species has both its P-P and P-W sliding
friction coefficients larger than the other (or if either one of these coefficients
is the same for both species) then segregation occurs. If, on the other hand, the
friction coefficients are the same, no axial segregation occurs even for particles
of different sizes; indeed an initially segregated system of this type, run \#W,
mixes (see later), so that at least some difference in frictional properties is
essential.

\begin{figure}
\includegraphics[scale=1.1]{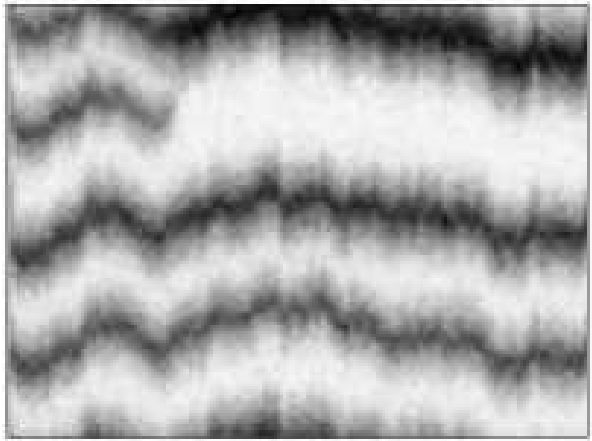}
\caption{\label{fig:stp}Space-time plot (\#P).}
\end{figure}

Run \#P in Fig.~\ref{fig:stp} shows what happens when more particles are added, in
this case almost half-filling the cylinder. Segregation still occurs, although the
band edges are starting to blur, reflecting reduced sharpness of the interfaces
between bands. Further filling, run \#Q, reduces the thoroughness of the
segregation even more and increases the time required for the bands to appear.

\subsection{Segregation in capped containers}

The periodic systems treated so far are useful for eliminating the potential role
of end caps as nucleation regions for segregation. However, periodic systems tend
to lack longitudinal stability, as a result of which the entire band pattern is
free to translate along the axis, making detailed analysis of the interface regions
between bands more difficult. The next series of runs demonstrates the kind of
segregation effects that can be seen when end caps are in place; the frictional
coefficients for the end walls are the same as for the curved wall.

\begin{figure}
\includegraphics[scale=1.1]{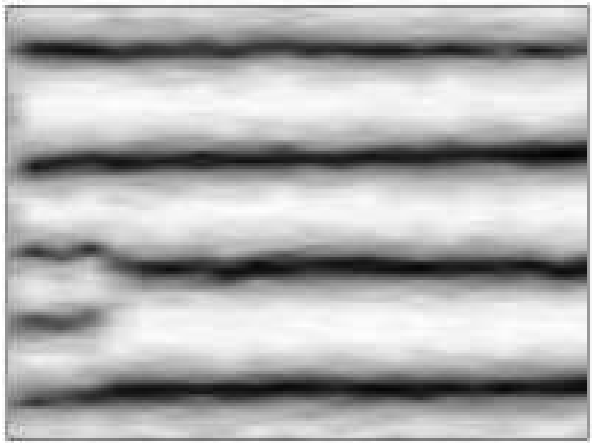}
\caption{\label{fig:sts}Space-time plot (\#R).}
\end{figure}

\begin{figure}
\includegraphics[scale=1.1]{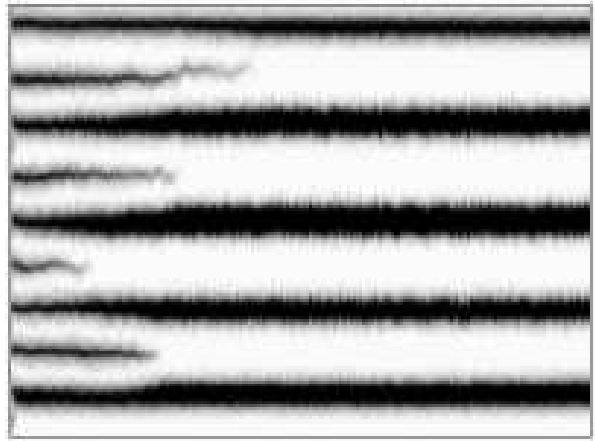}
\caption{\label{fig:stt}Space-time plot (\#S). }
\end{figure}

The first example, \#R, is shown in Fig.~\ref{fig:sts}. This run is fairly short,
allowing the initial band formation to be resolved; subsequently, one pair of dark
bands merges. What is different here is the absence of random movement of the band
pattern as a whole. The total number of bands has an odd value (in the periodic
systems it was of course always even), and it is the big particles with the larger
friction coefficients that occupy the end bands. The second example \#S, shown in
Fig.~\ref{fig:stt}, is for a longer cylinder. Here light bands are seen merging at
different times and there is an accompanying broadening of the remaining dark
bands; the bands corresponding to the big particles are wider, and again there is
no overall motion of the band pattern. Run \#T is an example of a system with
double the diameter; as in other cases with a larger interior cross-section, the
bands are not as sharp.

\begin{figure}
\includegraphics[scale=1.1]{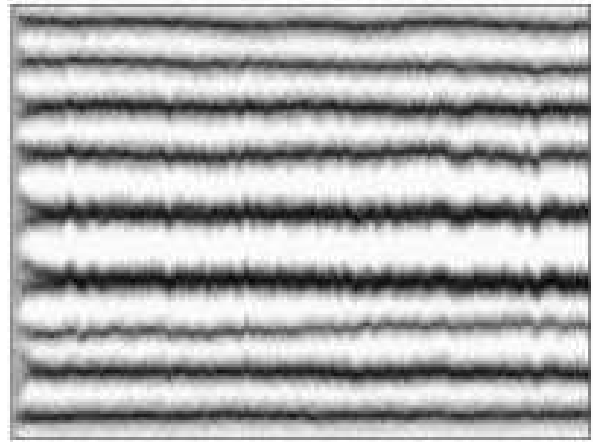}
\caption{\label{fig:stv}Space-time plot (\#U).}
\end{figure}

The lower $\Omega$ of run \#U, shown in Fig.~\ref{fig:stv}, produces an even richer
band structure. This is not such a long run (in terms of the number of rotations,
although not in terms of timesteps), and there is a hint that further band merging
might have occurred if the run had been longer. Overall, apart from stabilizing the
band pattern, there is no evidence to suggest that the end caps alter the behavior
in any way, and since band formation begins almost simultaneously at several
distinct locations, the ends do not function as nucleation sites.

\subsection{Absence of mixing in segregated systems}

The behavior of systems that are initially segregated is important for two reasons.
One is to establish whether the conditions under which the initially segregated
state persists or disappears are consistent with the previous conclusions based on
mixed systems. The other is to allow examination of whether, if the system showed
evidence of a preferred segregated state organized differently from the initial
state, it is capable of transforming itself from one state to another.

\begin{figure}
\includegraphics[scale=1.1]{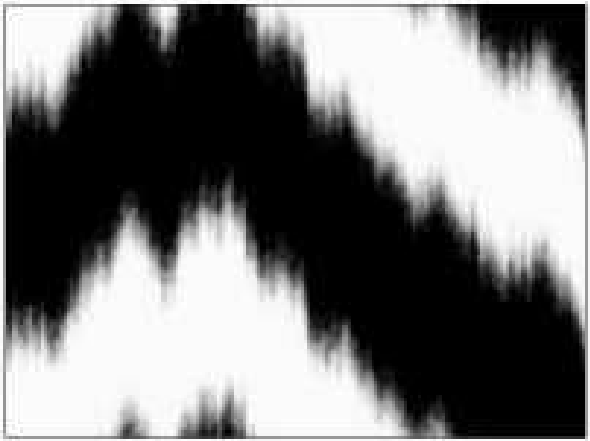}
\caption{\label{fig:stw}Space-time plot (\#V).}
\end{figure}

The first example, \#V, involves a system similar to \#A with periodic boundaries,
but now with an initial state in which all the small particles are located in a
band filling the central region of the cylinder. As shown in Fig.~\ref{fig:stw},
this segregated band persists throughout the run, and only its position changes.
Other sets of friction coefficients, corresponding to values for which segregation
was found to occur, show similar behavior, and for friction coefficients that do
not produce segregation, such as in \#W, the initial bands rapidly vanish as the
species mix. Since the periodic boundaries permit excessive motion of the pattern
as a whole, this run was repeated using a capped cylinder, run \#X, with the
initially segregated regions each occupying half the cylinder. Due to the
stabilizing influence of the end caps the pattern is completely stationary; again
there is no hint of mixing, and particles that do escape across the interface
promptly return to their own region. The evidence suggests that if the bands are
sharply delineated there is no tendency for any change to occur in the pattern, and
the multiple-band states that developed in initially mixed systems, such as \#S and
\#U, are not necessarily accessible to systems that are initially segregated;
memory of the initial state can apparently influence the long-term outcome of the
pattern evolution process.

\begin{figure}
\includegraphics[scale=1.1]{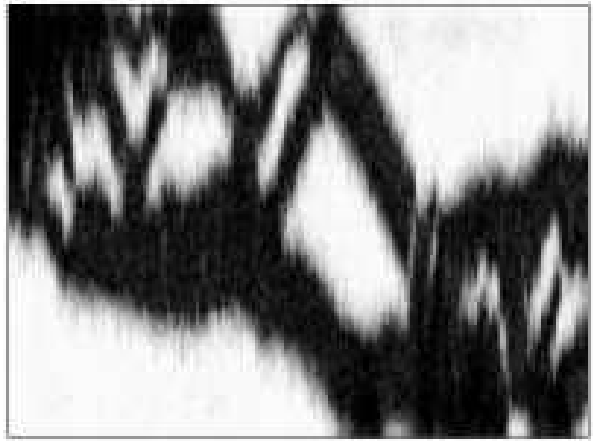}
\caption{\label{fig:stz}Space-time plot (\#Y).}
\end{figure}

\begin{figure}
\includegraphics[scale=0.58]{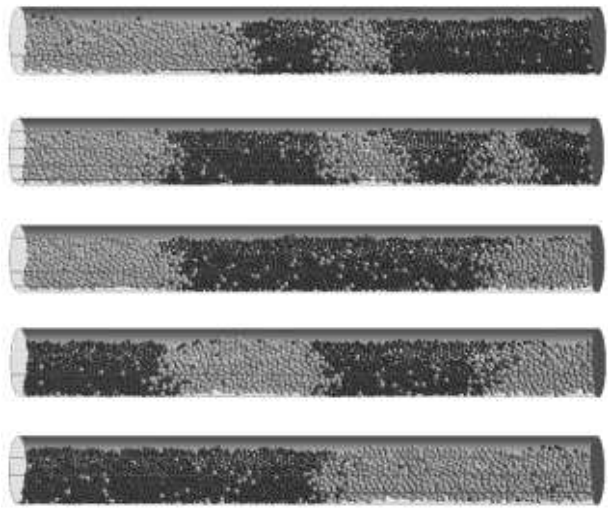}
\caption{\label{fig:vuz}Views of run \#Y at different times.}
\end{figure}

Modification of the friction coefficients makes it possible to reduce the stability
of the initial band pattern. Band splitting occurs as a result, and an example of
what happens, run \#Y (also with end caps), is shown in Fig.~\ref{fig:stz}. The
initial state has a single band boundary at the center of the cylinder, as in \#X,
and the run duration has been increased considerably to allow long-term behavior to
be monitored. The appearance of a new light region in the space-time plot
corresponds to the emergence of a band of big particles in the midst of a
small-particle region, and its disappearance corresponds to the vanishing of the
band (there are no new small-particle bands that form during this run). Multiple
band formation and disappearance events occur, reminiscent of Ref.~\cite{fre97} to
some extent, and at one point in the run the segregated regions can be seen to have
exchanged positions relative to the initial state; whether this system ever settles
into a steady state is a matter for speculation. 

In the corresponding initially mixed case, run \#E (a relatively short run), the
system segregated into eight bands; in \#Y there are intermediate states consisting
of four and six bands. The space-time plot is unable to convey an adequate
impression of this fascinating behavior, which may be analogous in some sense to
the waves observed experimentally \cite{cho97,cho98}. (A rough estimate of the
speed at which the bands move during the central portion of the run can be derived
from the gradient of the dark segment just to the right of the middle of
Fig.~\ref{fig:stz}. The result is approximately 1\% of the small particle diameter
per revolution; this is an order of magnitude less than in Ref.~\cite{cho98}
although it is not obvious that the effects are the same.) Figure~\ref{fig:vuz}
shows several snapshots from the run, including the reversed band state. Runs such
as this, with complex temporal evolution, are extremely sensitive to the initial
conditions, and when repeated using modified starting states are likely to exhibit
different, though (probably) statistically similar behavior; even those runs that
resulted in steady segregation patterns can have initial transient states that
depend on the initial conditions.

\section{Conclusions}

The simulations described in this paper have displayed a variety of phenomena that
largely correspond to what is observed experimentally. The formation of axially
segregated bands is the most basic of the effects, and the fact that segregation
occurs under a variety of conditions is a measure of the success of the model. More
complex aspects of the behavior have also been addressed briefly, including
behavior in the interior, traveling patterns, and memory effects. Even if not all
the parameter combinations examined correspond to physically realizable systems,
the fascinating pattern development observed here merits examination in its own
right, since corresponding effects do appear in real granular systems. No attempt
has been made to probe the actual mechanisms involved, leaving this aspect of the
problem for future detailed study; clearly, as is evident from actually watching
the simulations in progress, both avalanches and diffusion play a significant role
in much of what occurs, but it has yet to be established whether these processes
alone are sufficient to account for all aspects of the behavior.

Although the simulations have succeeded in qualitatively reproducing features
observed experimentally it is important to reiterate possible discrepancies, the
resolution of which may help in understanding the successes and limitations of the
models used for granular simulation, as well as suggesting where additional
experiments could be helpful. Two such issues were mentioned previously; the first
is the relative roughness of the big and small particles, and which species should
have the higher friction coefficients, a subject on which experiment
\cite{zik94,hil97f} has yet to provide an unequivocal answer; the second is the
question of which particle species tends to congregate near the cylinder axis,
since the MRI experiments were able to show the locations of just one of the
species \cite{hil97s,hil97f}. In addition, as is the case with other granular
simulations, the degree to which the omission of static friction from the model
adversely affects the results can only be determined empirically. Hopefully,
further experimental and simulational effort will help resolve these issues.

It must be emphasized that the conclusions concerning the conditions under which
segregation occurs in the model apply only to the specific parameter values
considered here; crossover from segregating to non-segregating behavior occurs at
intermediate parameter settings, and more extensive exploration of the phase
diagram is needed to fill in the gaps. Furthermore, because some slowly-evolving
aspects of the behavior can be seen only over comparatively long time intervals,
the question of whether the patterns observed here represent true final states, or
whether the bands are long-lived but only metastable (in other words, whether or
not the system is in dynamic equilibrium) also requires further study, although
often the band patterns and the actual particle dynamics provided no hint that
further changes ought to be anticipated.

The analysis has focused on overall axial segregation, but there are many other
aspects that await future investigation, some of which are likely to entail more
extensive computations than employed here. Beyond what can be seen in some of the
images showing internal particle organization, the issue of radial behavior,
especially radial segregation, has not been addressed; cylinders with a larger
diameter may be needed to properly analyze this behavior. In order to learn more
about wavelength selection in the segregation bands, if such preferences really
exist, repeated long runs with larger systems are needed. The full
three-dimensional organization of the particle flows can also be studied; this
would provide information on the way interfaces form and decay, the nature of the
internal axial and radial flows, and the different kinds of behavior occurring in
regions near the center of the layer and close to the free surface; the ability of
simulation to probe the interior, in a conceptually similar manner to MRI, is a
major advantage of this kind of modeling approach.

\begin{acknowledgments}
This work was partially supported by the Israel Science Foundation.
\end{acknowledgments}

\bibliography{granrot}

\end{document}